\documentclass[twocolumn,english,aps,pra,showpacs]{revtex4}
\usepackage[T1]{fontenc}
\usepackage[latin1]{inputenc}
\usepackage{amsmath}
\usepackage{amssymb}
\usepackage{graphicx}
\usepackage{esint}

\makeatletter
\@ifundefined{textcolor}{}
{%
 \definecolor{BLACK}{gray}{0}
 \definecolor{WHITE}{gray}{1}
 \definecolor{RED}{rgb}{1,0,0}
 \definecolor{GREEN}{rgb}{0,1,0}
 \definecolor{BLUE}{rgb}{0,0,1}
 \definecolor{CYAN}{cmyk}{1,0,0,0}
 \definecolor{MAGENTA}{cmyk}{0,1,0,0}
 \definecolor{YELLOW}{cmyk}{0,0,1,0}
}

\@ifundefined{definecolor}{\@ifundefined{definecolor}{\@ifundefined{definecolor}{\@ifundefined{definecolor}
 {\usepackage{color}}{}
}{}}{}}{}\makeatother

\makeatother

\usepackage{babel}

\makeatother

\usepackage{babel}

\makeatother

\usepackage{babel}

\makeatother

\usepackage{babel}

\makeatother

\usepackage{babel}

\makeatother

\usepackage{babel}
\begin{document}

\title{Probing Majorana fermions in spin-orbit coupled atomic Fermi gases}

\author{Xia-Ji Liu$^{1}$, Lei Jiang$^{2}$, Han Pu$^{2}$, and Hui Hu$^{1}$}

\affiliation{$^{1}$ACQAO and Centre for Atom Optics and Ultrafast Spectroscopy,
Swinburne University of Technology, Melbourne 3122, Australia \\
 $^{2}$Department of Physics and Astronomy, and Rice Quantum Institute,
Rice University, Houston, TX 77251, USA}

\date{\today}
\begin{abstract}
We examine theoretically the visualization of Majorana fermions in
a two-dimensional trapped ultracold atomic Fermi gas with spin-orbit
coupling. By increasing an external Zeeman field, the trapped gas
transits from non-topological to topological superfluid, via a mixed
phase in which both types of superfluids coexist. We show that the
zero-energy Majorana fermion, supported by the topological superfluid
and localized at the vortex core, may be visible through (i) the core
density and (ii) the local density of states, which are readily measurable
in experiment. We present a realistic estimate on experimental parameters
for ultracold $^{40}$K atoms. 
\end{abstract}


\maketitle
Majorana fermion \cite{Majorana} - particle that is its own antiparticle
- has attracted considerable attentions from a wide area of physics
\cite{WilczekPerspective}. A particular interest comes from its non-Abelian
exchange statistics that is crucial for topological quantum computation
\cite{KitaevAnnPhys2006,NayakRMP2008}. Two well separated Majorana
fermions may form a non-local fermionic state, as a non-local qubit
for inherently fault-tolerant quantum memory. As a portal for future
quantum technology, the realization of Majorana fermions in a highly
controllable manner is of great importance and a timely quest.

Majorana fermions are believed to exist in a number of two-dimensional
(2D) physical settings, including fractional quantum Hall states at
filling $\nu=5/2$ \cite{MooreNPB1991}, vortex states of $p_{x}+ip_{y}$
superconductors/superfluids \cite{ReadPRB2000,MizushimaPRL2008},
and surfaces of 3D topological insulators in proximity to a $s$-wave
superconductor \cite{FuPRL2008}. Majorana fermions may also emerge
in 1D quantum systems with strong spin-orbit coupling, such as quantum
wires \cite{LutchynPRL2010} and optically trapped 1D fermionic atoms
\cite{JiangPRL2011}. All these appealing proposals are yet to be
realized experimentally.

In this work, we examine the possibility of observing Majorana fermions
in the vortex core of a spin-orbit coupled ultracold atomic Fermi
gas in 2D harmonic traps. This is a scenario discussed earlier by
several researchers \cite{ZhangPRL2008,SatoPRL2009,ZhuPRL2011}, based
on a theoretical concept originated from Jackiw and Rossi, who predicted
the existence of vortex-core Majorana fermions in (2+1) dimensional
Dirac theory \cite{JackiwNPB1981}. Here, we perform a fully microscopic
calculation with Bogoliubov-de Gennes (BdG) equation, which enables
simulations with realistic experimental parameters. Our study is motivated
by the recent creation of non-Abelian gauge fields in a Bose-Einstein
condensate (BEC) of $^{87}$Rb atoms \cite{LinNature2011} and its possible
realization in fermionic $^{40}$K atoms \cite{SauPRB2011}.

We find that by increasing a Zeeman field, a topological superfluid
emerges from the trap edge and extends gradually to the whole Fermi
cloud, supporting zero-energy states (ZES) at the vortex core and
trap edge. This topological phase transition is detectable through
a sudden change of the atomic density inside the vortex core, associated
with the occupation of the Majorana state. We show that the wave function
of the Majorana fermions can be inferred from the local density of
states (LDOS) at the core.

\textit{Mean-field BdG equation. --- }We consider a trapped 2D atomic
Fermi gas subject to Rashba spin-orbit coupling $V_{{\rm SO}}(\mathbf{{\bf r}})=-i\lambda(\partial_{y}+i\partial_{x})$
and a Zeeman field $h$, which may be prepared in a single pancake-like
optical trap $V({\bf r},z)=M[\omega_{\perp}^{2}r^{2}+\omega_{z}^{2}z^{2}]/2$
with trapping frequencies $\omega_{z}\gg\omega_{\perp}$. We note
that 2D Fermi gas has recently been realized in experiments \cite{DykePRL2011,FrohlichPRL2011}.
The system is described by ${\cal H}=\int d{\bf r}\,[{\cal H}_{0}({\bf r})+{\cal H}_{I}({\bf r})]$,
where 
\begin{equation}
{\cal H}_{0}({\bf r})={\displaystyle \sum_{\sigma=\uparrow,\downarrow}}\psi_{\sigma}^{\dagger}{\cal H}_{\sigma}^{S}({\bf r})\psi_{\sigma}+\left[\psi_{\uparrow}^{\dagger}V_{{\rm SO}}({\bf r})\psi_{\downarrow}+\text{H.c.}\right]
\end{equation}
 and ${\cal H}_{I}({\bf r})=U_{0}\psi_{\uparrow}^{\dagger}({\bf r})\psi_{\downarrow}^{\dagger}({\bf r})\psi_{\downarrow}({\bf r})\psi_{\uparrow}({\bf r})$
describes the contact interaction between opposite spins. Here $\psi_{\uparrow,\downarrow}^{\dagger}$
are the creation field operators for the spin-up and -down atoms,
${\cal H}_{\sigma}^{S}=-\hbar^{2}\nabla^{2}/(2M)+M\omega_{\perp}^{2}r^{2}/2-\mu-h\sigma_{z}$
is the single-particle Hamiltonian in reference to the chemical potential
$\mu$. The interaction strength $U_{0}$ is to be regularized via
$1/U_{0}+\sum_{{\bf k}}1/(\hbar^{2}{\bf k}^{2}/M+E_{a})=M/(4\pi\hbar^{2})\ln(E_{a}/E)$.
Here $E_{a}$ is the binding energy of the two-body bound state \cite{RanderiaPRL1989,PetrovPRA2003}
and $E>0$ is the relative collision energy.

\begin{figure*}[htp]
\begin{centering}
\includegraphics[clip,width=0.75\textwidth]{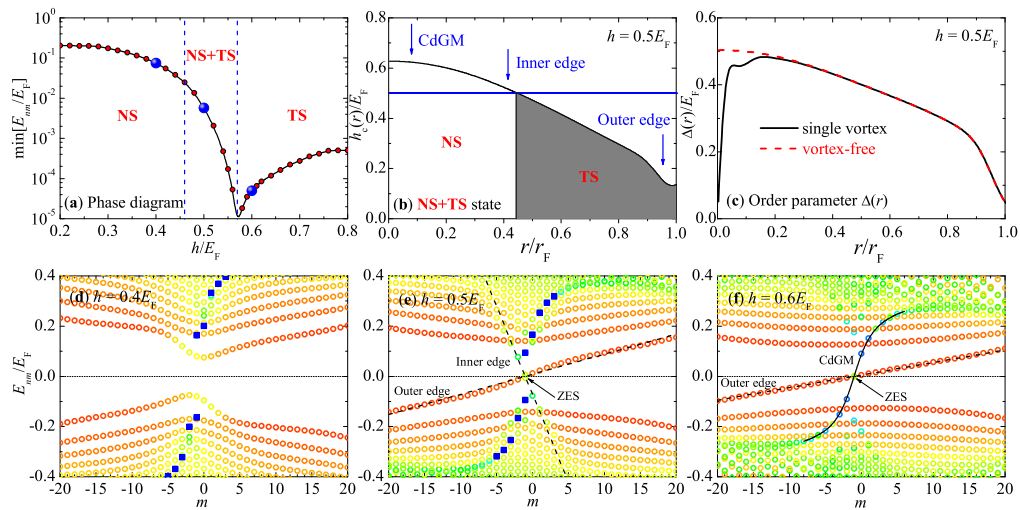} 
\par\end{centering}

\caption{(color online) (a) Phase diagram, along with the lowest eigenenergy
of Bogoliubov spectrum. (b) Configuration of the mixed phase at $h=0.5E_{F}$.
The local critical field is calculated in the vortex-free state. (c)
The gap profile with and without vortex at $h=0.5E_{F}$. (d), (e),
and (f) Energy spectrum at $h/E_{F}=0.4$, $0.5$, and $0.6$ (where
$E_{F}=\hbar^{2}k_{F}^{2}/(2M)=\sqrt{N}\hbar\omega_{\perp}$ is the
Fermi energy) in the presence of a single vortex. The color of symbols
indicates the mean radius $\sqrt{\left\langle r^{2}\right\rangle }/r_{F}$
(where $r_{F}=(4N)^{1/4}\sqrt{\hbar/(M\omega_{\perp})}$ is the Fermi
radius) of the eigenstate, which is defined by $\left\langle r^{2}\right\rangle =\int r^{2}[\left|u_{\uparrow}\right|^{2}+\left|u_{\downarrow}\right|^{2}+\left|v_{\uparrow}\right|^{2}+\left|v_{\downarrow}\right|^{2}]d{\bf r}$.
The color of symbols changes from blue when the excited state is localized
at the trap center to red when its mean radius approaches the Thomas-Fermi
radius. The CdGM states are indicated by blue squares in (d) and (e),
and by a solid line in (f). The dashed lines show the inner edge and
outer edge branches. }

\label{fig1} 
\end{figure*}

The low-energy fermionic quasiparticles are solved by the mean-field
BdG approach, ${\cal H}_{{\rm BdG}}\Psi_{\eta}\left({\bf r}\right)=E_{\eta}\Psi_{\eta}\left({\bf r}\right)$.
Using the convention for Nambu spinors $\Psi_{\eta}\left({\bf r}\right)=[u_{\uparrow\eta},u_{\downarrow\eta},v_{\uparrow\eta},v_{\downarrow\eta}]^{T}$,
the BdG Hamiltonian reads, 
\begin{equation}
{\cal H}_{{\rm BdG}}=\left[\begin{array}{cccc}
{\cal H}_{\uparrow}^{S}({\bf r}) & V_{{\rm SO}}({\bf r}) & 0 & -\Delta({\bf r})\\
V_{{\rm SO}}^{\dagger}({\bf r}) & {\cal H}_{\downarrow}^{S}({\bf r}) & \Delta({\bf r}) & 0\\
0 & \Delta^{*}({\bf r}) & -{\cal H}_{\uparrow}^{S}({\bf r}) & V_{{\rm SO}}^{\dagger}({\bf r})\\
-\Delta^{*}({\bf r}) & 0 & V_{{\rm SO}}({\bf r}) & -{\cal H}_{\downarrow}^{S}({\bf r})
\end{array}\right],
\end{equation}
 where $\Delta=-(U_{0}/2)\sum_{\eta}[u_{\uparrow\eta}v_{\downarrow\eta}^{*}f(E_{\eta})+u_{\downarrow\eta}v_{\uparrow\eta}^{*}f(-E_{\eta})]$
is the order parameter, to be solved self-consistently in conjunction
with the atomic densities, $n_{\sigma}\left({\bf r}\right)=(1/2)\sum_{\eta}[\left|u_{\sigma\eta}\right|^{2}f(E_{\eta})+\left|v_{\sigma\eta}\right|^{2}f(-E_{\eta})]$.
Here $f\left(x\right)\equiv1/\left(e^{x/k_{B}T}+1\right)$ is the
Fermi distribution function. The chemical potential $\mu$ is determined
by the total atom number $N=\int d{\bf r[}n_{\uparrow}\left({\bf r}\right)+n_{\downarrow}\left({\bf r}\right)]$.
With a single vortex at trap center, we take $\Delta({\bf r})=\Delta(r)e^{-i\varphi}$
and decouple the BdG equation into different angular momentum channels
indexed by an integer $m$. The quasiparticle wave functions take
the form, $[u_{\uparrow\eta}(r)e^{-i\varphi},u_{\downarrow\eta}(r),v_{\uparrow\eta}(r)e^{i\varphi},v_{\downarrow\eta}(r)]e^{i(m+1)\varphi}/\sqrt{2\pi}$.
We have solved self-consistently the BdG equations using the basis
expansion method. For the results presented here, we have taken $N=400$
and $T=0$. We have used $E_{a}=0.2E_{F}$ and $\lambda k_{F}/E_{F}=1$,
which are typical parameters that can be readily realized in a 2D
$^{40}$K Fermi gas \cite{FrohlichPRL2011}. Other sets of parameters,
with varying interaction strength, SO coupling strength and temperature,
have also been tried.

The use of Nambu spinor representation leads to an inherent redundancy
built into the BdG Hamiltonian. ${\cal H}_{{\rm BdG}}$ is invariant
under the {\em particle-hole} transformation, $u_{\sigma}\left({\bf r}\right)\rightarrow v_{\sigma}^{*}\left({\bf r}\right)$
and $E_{\eta}\rightarrow-E_{\eta}$. Thus, every eigenstate with energy
$E$ has a partner at $-E$. These two states describe the same physical
degrees of freedom, as the Bogoliubov quasiparticle operators associated
with them satisfy $\Gamma_{E}=\Gamma_{-E}^{\dagger}$. In the expressions
for order parameter and atomic density, this redundancy has been removed
by multiplying a factor of $1/2$.

\textit{Majorana fermions}. --- The particle-hole redundancy, however,
is very useful to illustrate a non-trivial feature when the Zeeman
field $h$ is beyond a threshold, $h>h_{c}=\sqrt{\mu^{2}+\Delta^{2}}$
and the system is in a topological state \cite{FuPRL2008}, hosting
ZES within the energy gap. Due to $E=0$, the associated quasiparticle
operators satisfy $\Gamma_{0}=\Gamma_{0}^{\dagger}$. Thus, a zero-energy
quasiparticle is its own antiparticle - exactly the defining feature
of a Majorana fermion \cite{WilczekPerspective}. Because of the redundant
particle-hole representation, the ZES or Majorana fermion is a {\em
half} of ordinary fermion and thus must always come in pairs. Each
of the paired states, localized \emph{separately} in real space, can
hardly be pushed away from $E=0$ by a local perturbation \cite{NayakRMP2008,ReadPRB2000},
giving rise to the intrinsic topological stability enjoyed by Majorana
fermions. It is straightforward to show from the BdG Hamiltonian that
the wave function of Majorana fermions should satisfy either $u_{\sigma}\left({\bf r}\right)=v_{\sigma}^{*}\left({\bf r}\right)$
or $u_{\sigma}\left({\bf r}\right)=-v_{\sigma}^{*}\left({\bf r}\right)$.
The former follows directly from the particle-hole symmetry. For the
latter, the related quasiparticle operator satisfies $\tilde{\Gamma}_{0}=-\tilde{\Gamma}_{0}^{\dagger}$.
This is necessary for expressing an orindary fermion using paired
Majorana fermions \cite{footnote}.

\textit{Phase diagram}. --- Figure \ref{fig1} reports the phase diagram
(a) along with the quasiparticle energy spectrum of different phases
(d, e, and f) in the presence of a single vortex. By increasing the
Zeeman field, the system evolves from a non-topological state (NS)
to a topological state (TS), through an intermediate mixed phase in
which NS and TS coexist. The mixed phase, unique for a trapped system,
can be easily understood from the point of view of local density approximation,
in which the local chemical potential $\mu({\bf r})=\mu-M\omega_{\perp}^{2}r^{2}/2$
and the order parameter $\Delta({\bf r})$ decrease continuously away
from the trap center. As a result, the local critical Zeeman field
$h_{c}({\bf r})=\sqrt{\mu^{2}({\bf r})+\Delta^{2}({\bf r})}$ becomes
smaller at the trap edge, as shown in Fig. (1b) for $h=0.5E_{F}$.
This creates a ring of TS at the outer region where $h>h_{c}({\bf r})$,
surrounding the inner region which is non-topological. The full topological
transition occurs when the Zeeman field is larger than the critical
field at the trap center, $h>\sqrt{\mu^{2}+\Delta^{2}(0)}$, where
$\Delta(0)$ is the gap at trap center in the absence of the vortex.

The topological phase transition into TS is well characterized by
the low-lying quasiparticle spectrum, which has the particle-hole
symmetry $E_{m+1}=-E_{-(m+1)}$. As shown in Fig. (1d), the spectrum
of the NS is gapped by order parameter, even at the trap edge. In
the mixed phase (Fig. (1e)), however, three branches with small energy
spacing appear. The branches labeled as ``Outer edge'' and ``Inner
edge'' consist of the eigenstates with wave functions localized at
the edges of the TS as Andreev bound states, as indicated by the arrows
in Fig. (1b). The two edge states at $m=-1$ have nearly zero energy.
The other branch, shown by blue squares, is a series of discrete localized
states at the vortex core, i.e., the so--called Caroli-de Gennes-Matricon
(CdGM) states \cite{CdGM}. When the TS extends over the whole cloud
(Fig. (1f)), the dispersion of the inner edge branch moves into the
continuum. The energy of the state with $m=-1$ in both the CdGM branch
and outer edge branch becomes essentially zero.

\begin{figure}[htp]
\begin{centering}
\includegraphics[clip,width=0.4\textwidth]{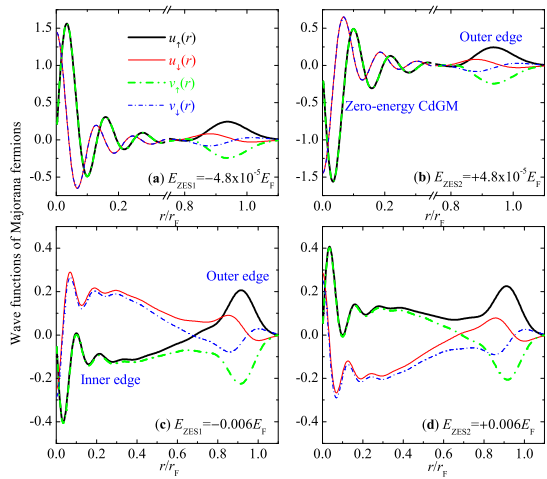} 
\par\end{centering}

\caption{(color online) Wave functions of the two lowest-energy modes in the
full topological state at $h=0.6E_{F}$, (a) and (b), and in the partial
topological state at $h=0.5E_{F}$, (c) and (d). The wave functions
are in units of $\sqrt{M\omega_{\perp}/\hbar}$ .}

\label{fig2} 
\end{figure}

The eigenstates with nearly zero energy at $m=-1$, i.e., the two
edge states in the mixed phase as well as the outer edge state and
CdGM state in the full TS, can be identified as zero-energy Majoranan
fermions in the thermodynamic limit. To show this, we plot in Fig.~\ref{fig2}
the wave function of these eigenstates in the full TS (a and b) and
in the mixed phase (c and d). In both cases, we observe a bond and
anti-bond hybridization between two well-localized wave functions:
one satisfies $u_{\sigma}\left({\bf r}\right)=v_{\sigma}^{*}\left({\bf r}\right)$
and the other $u_{\sigma}\left({\bf r}\right)=-v_{\sigma}^{*}\left({\bf r}\right)$,
which is exactly the symmetry of the wave function required by Majorana
fermions. The hybridization is caused by a quasiparticle tunneling
between paired Majorana states \cite{MizushimaPRA2010}, which leads
to the splitting of degenerate zero-energy states to finite energies
$\pm E_{{\rm ZES}}$. The tunneling barrier between the two edge states
is lower, so the energy splitting is relatively larger (i.e., $E_{{\rm ZES}}\approx0.006E_{F}$).
In contrast, the tunneling between the outer edge state and CdGM state
seems to be more difficult, giving rise to an exponentially small
splitting (i.e., $E_{{\rm ZES}}\approx4.8\times10^{-5}E_{F}$).

The different phases can therefore be identified from the lowest eigenenergy
of energy spectrum, as plotted in Fig. (1a). Within the NS, it decreases
slowly as the Zeeman field increases. The decrease becomes exponentially
fast with the appearance of a partial TS and a minimum is reached
when the cloud just becomes fully TS. Further increase of the Zeeman
field will reduce the order parameter and hence the tunneling barrier
between the two Majorana states, which are localized respectively
at the vortex core and trap edge, leading to a steady increase of
the lowest eigenenergy. We note that the exponentially small energy
of Majorana fermions in the full TS, inherent to the finiteness of
trapped Fermi cloud, should be suppressed by increasing the number
of total atoms.

\textit{Probing Majorana fermions.} --- In the TS, the occupation
of the Majorana vortex-core state affects significantly the atomic
density and LDOS of the Fermi cloud near the trap center, which in
turn gives an unambiguous experimental signature for observing Majorana
fermions.

\begin{figure}[htp]
\begin{centering}
\includegraphics[clip,width=0.35\textwidth]{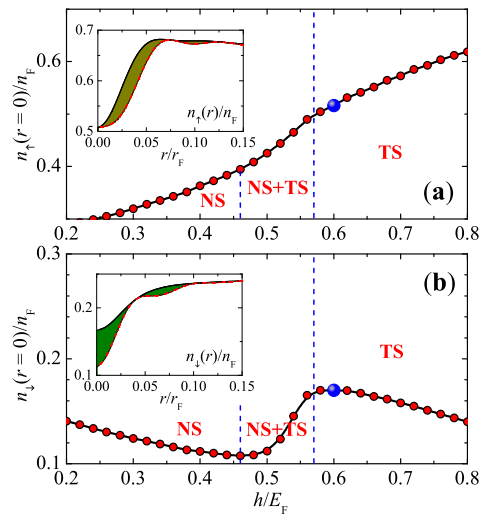} 
\par\end{centering}

\caption{(color online) Zeeman field dependence of spin-up (a) and spin-down
(b) densities at the vortex core. The density is normalized by the
Thomas-Fermi density $n_{F}=(\sqrt{N}/\pi)\sqrt{M\omega_{\perp}/\hbar}$.
The insets show the core density distributions at $h=0.6E_{F}$. The
red dot-dashed lines show the result by excluding artificially the
Majorana vortex core state, whose contribution is shown by the shaded
area.}

\label{fig3} 
\end{figure}

Figure \ref{fig3} presents the spin-up and -down densities at the
trap center, $n_{\uparrow}(0)$ and $n_{\downarrow}(0)$, as a function
of the Zeeman field. In general, $n_{\uparrow}(0)$ and $n_{\downarrow}(0)$
increases and decreases respectively with increasing field. However,
we find a sharp increase of $n_{\downarrow}(0)$ when the system evolves
from the mixed phase to the full TS. Accordingly, a change of slope
or kink appears in $n_{\uparrow}(0)$. The increase of $n_{\downarrow}(0)$
is associated with the {\em gradual} formation of the Majorana
vortex-core mode, whose occupation contributes notably to atomic density
due to the {\em large} amplitude of its localized wave function.
We plot in the inset of Fig. (3b) $n_{\downarrow}(0)$ at $h=0.6E_{F}$,
with or without the contribution of the Majorana mode, which is highlighted
by the shaded area. This contribution is apparently absent in the
NS. Thus, a sharp increase of $n_{\downarrow}(0)$, detectable in
in-situ absorption imaging, signals the topological phase transition
and the appearance of the Majorana vortex-core mode. This feature
persists at typical experimental temperature, i.e., $T=0.1T_{F}$.
We note that experimentally it is more favorable to take a time-of-flight
imaging of the cloud after an expansion time, in order to have enough
resolution to visualize the vortex core. We anticipate the sharp increase
in $n_{\downarrow}(0)$ may persist for a short expansion time. Alternatively, 
we may tune quickly an external magnetic field to take the Fermi system to the BEC limit across 
Feshbach resonances. In this way, the vortex core can be imaged clearly after the time-of-flight 
just like in an atomic BEC.

\begin{figure}[htp]
\begin{centering}
\includegraphics[clip,width=0.4\textwidth]{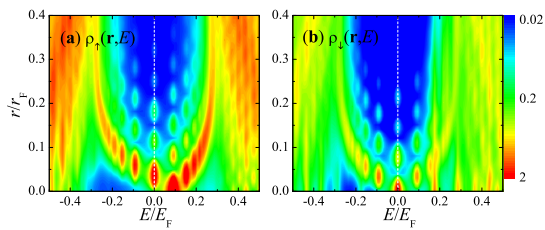} 
\par\end{centering}

\caption{(color online) Log-scale contour plot of the LDOS for spin-up (a)
and spin-down (b) atoms at $h=0.6E_{F}$. Here we use $\Gamma=0.01E_{F}$.
The LDOS is units of $n_{F}/E_{F}$.}

\label{fig4} 
\end{figure}

In the full TS, the wave function of the Majorana mode can be determined
by measuring LDOS through spatially resolved rf-spectroscopy \cite{ShinPRL2007,JiangPRA2011},
which provides a cold-atom analog of the widely used scanning tunneling
microscope in solid state. We show in Fig.~\ref{fig4} the spin-up
and -down LDOS at $h=0.6E_{F}$ defined as $\rho_{\sigma}({\bf r},E)=(1/2)\sum_{\eta}[\left|u_{\sigma\eta}\right|^{2}\delta(E-E_{\eta})+\left|v_{\sigma\eta}\right|^{2}\delta(E+E_{\eta})]$,
where the $\delta$-function can be simulated by a Lorentzian distribution
with a suitable energy broadening $\Gamma$. Inside the vortex core,
the contribution from the Majorana mode and other CdGM states is clearly
visible within the superfluid gap. In the case of $\Gamma,k_{B}T<\Delta E$,
where $\Delta E\sim\Delta^{2}(0)/(2E_{F})$ is the energy spacing
of CdGM states \cite{CdGM} which for typical parameters as used in
our calculation turns out to be about 10 nK, the Majorana fermion
contribution $\rho_{\sigma,{\rm ZES}}({\bf r},0)$ may be singled
out. As $\rho_{\sigma,{\rm ZES}}({\bf r},0)\propto\left|u_{\sigma\eta}({\bf r})\right|^{2}=\left|v_{\sigma\eta}({\bf r})\right|^{2}$,
the spatially resolved rf-spectroscopy maps out directly the wave
function of the Majorana vortex-core state.

In closing, we note that 2D ultracold atomic Fermi gases are an ideal
platform for probing and manipulating Majorana fermions because of
the unprecedented controllability and flexibility. This is particularly
useful for the purpose of topological quantum computation, using Majorana
fermions as qubits \cite{NayakRMP2008}. For instance, two 2D atomic
Fermi gases formed by a double-well potential along $z$-axis, each
of which has a single vortex at the center, can host four Majorana
fermions for carrying out the basic information process. In this configuration,
inter-well quantum tunneling of Majorana fermions is possible, providing
another potential means to detect the interesting topological phase
transition.

\textit{Acknowledgment} --- We thank Peter Drummond, Chris Vale, and
Peter Hannaford for helpful discussions. HH and XJL are supported
by the ARC Discovery Project (Grant No. DP0984522 and DP0984637) and
NFRP-China (Grant No. 2011CB921502). HP is supported by the NSF and
the Welch Foundation (Grant No. C-1669).


\begin{thebibliography}{10}
\bibitem{footnote}The ordinary fermion operator at $E=0$ is given
by $c=\Gamma_{0}+\tilde{\Gamma}_{0}$. By defining Majorana operators
$\gamma_{1}=\Gamma_{0}$ and $\gamma_{2}=i\tilde{\Gamma}_{0}$, we
express $c=\gamma_{1}-i\gamma_{2}$, as anticipated.
\end{thebibliography}
\end{document}